\documentclass[aps,prb,citeautoscript,
 preprint, amsmath, amssymb]{revtex4-2}

\usepackage{bm}% bold math
\usepackage{amsmath}
\usepackage{graphicx}
\usepackage{dcolumn}
\usepackage{bm}
\usepackage{lipsum}
\usepackage{gensymb}
\usepackage{siunitx}
\usepackage{tabularx}
\usepackage{multirow}
\usepackage{setspace}

\usepackage{booktabs}
\usepackage{natbib}

\usepackage[utf8]{inputenc}
\usepackage[english]{babel}

\newcommand{\rpm}{\sbox0{$1$}\sbox2{$\scriptstyle\pm$}\raise\dimexpr(\ht0-\ht2)/2\relax\box2 }

\begin{document}

\preprint{APS/123-QED}

\title{Quantifying Atomic Structural Disorder\\ Using Procrustes Shape Analysis}

\author{Jinchen Han}
\affiliation{Department of Mechanical Engineering, Carnegie Mellon University, \\Pittsburgh, Pennsylvania 15213, USA}
\author{Henry T. Aller}
\affiliation{Department of Mechanical Engineering, Carnegie Mellon University, \\Pittsburgh, Pennsylvania 15213, USA}
\altaffiliation{Department of Materials Science and Engineering, Carnegie Mellon University, \\Pittsburgh, Pennsylvania 15213, USA}
\author{Alan J. H. McGaughey}
\affiliation{Department of Mechanical Engineering, Carnegie Mellon University, \\Pittsburgh, Pennsylvania 15213, USA}
\email{mcgaughey@cmu.edu}

\date{\today}

\maketitle

\section{System Preparation}

In this study, we added vacancies to change the structure of a Si/Ge interface. We applied Procrustes analysis\cite{Gower}  for disorder quantification and identifying different local atomic environments. 

We performed molecular dynamics simulations of a Si/Ge interface using the LAMMPS package \cite{LAMMPS}, shown in Figure 1. The atomic interactions between all atoms were described by the Tersoff potential\cite{tersoff1989modeling}. All simulations used a time step as 0.5 fs\cite{JAP_a/a}. The system contains 3$\times$3$\times$56 unit cells including 4032 atoms\cite{JAP_a/a}. Additional 2 unit-cell-thick layers of fixed atoms were set on opposite ends of the simulation in the $z$ direction to rebound moving atoms and set a non-periodic, adiabatic boundary. Periodic boundary conditions were applied in the $x$ and $y$ directions.

To generate vacacies and interfacial disorder, we deleted a small number of atoms near the interface according to a Gaussian probability function,
\begin{equation}
f(z)=A_{\rm m} \: {\rm exp}\left[-\frac{1}{2}\left(\frac{z-\mu}{\sigma}\right)^2\right],
\end{equation}
where $A_{\rm m}$ is the highest probability of producing vacancies at the interface, $\mu$ is the interface position and $\sigma$ is a length scale that controls the range of vacancies. Because of the high stability of the Si/Ge interface, we set the highest probability $A_{\rm m}$ as 0.3 to create a reasonable amount of vacancies.  By varying $\sigma$, we are able to alter how far from the interface vacancies are created. Four separate interfaces were constructed using different $\sigma$ values, with their targeted distributions plotted in Figure 2.

\begin{figure}
    \includegraphics[width=1\linewidth]{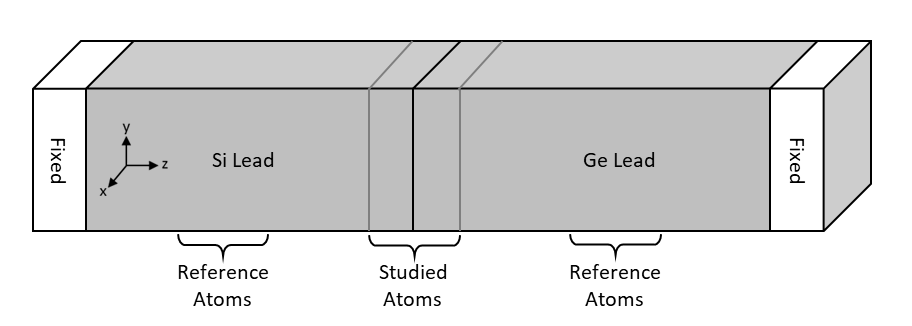}\par
    \caption{Schematic figure of the simulation setting for the Si/Ge interface.}
    \label{f-2} 
\end{figure}

Following vacancy creation, we relaxed the interfacial system. Both Si and Ge are in a diamond structure to maintain a coherent interface. 5 million time steps of anisotropic $NPT$ simulation were performed at zero pressure and a temperature of 300 K. The resulting cross-sectional lattice constant is 5.569 \AA, which is close to the average of bulk Si and Ge. This endows the Si and Ge with different stress states to simulate the impact of the interfacial lattice mismatch. With the fixed cross-sectional area, we then optimized the length of the systems in the $z$-direction. Specifically, we performed 2 million steps of $NVT$ simulation at a temperature of 300 K with multiple lengths to obtain a length with the minimum stress along the $z$-direction. 

Using the fully relaxed system, we performed 2 million steps of $NVT$ simulation to set a temperature of 300 K. Then, the simulation was maintained under the $NVE$ ensemble for 20 million steps. The averaged atomic positions were collected over another 80 million steps.

\begin{figure}
    \includegraphics[width=0.5\linewidth]{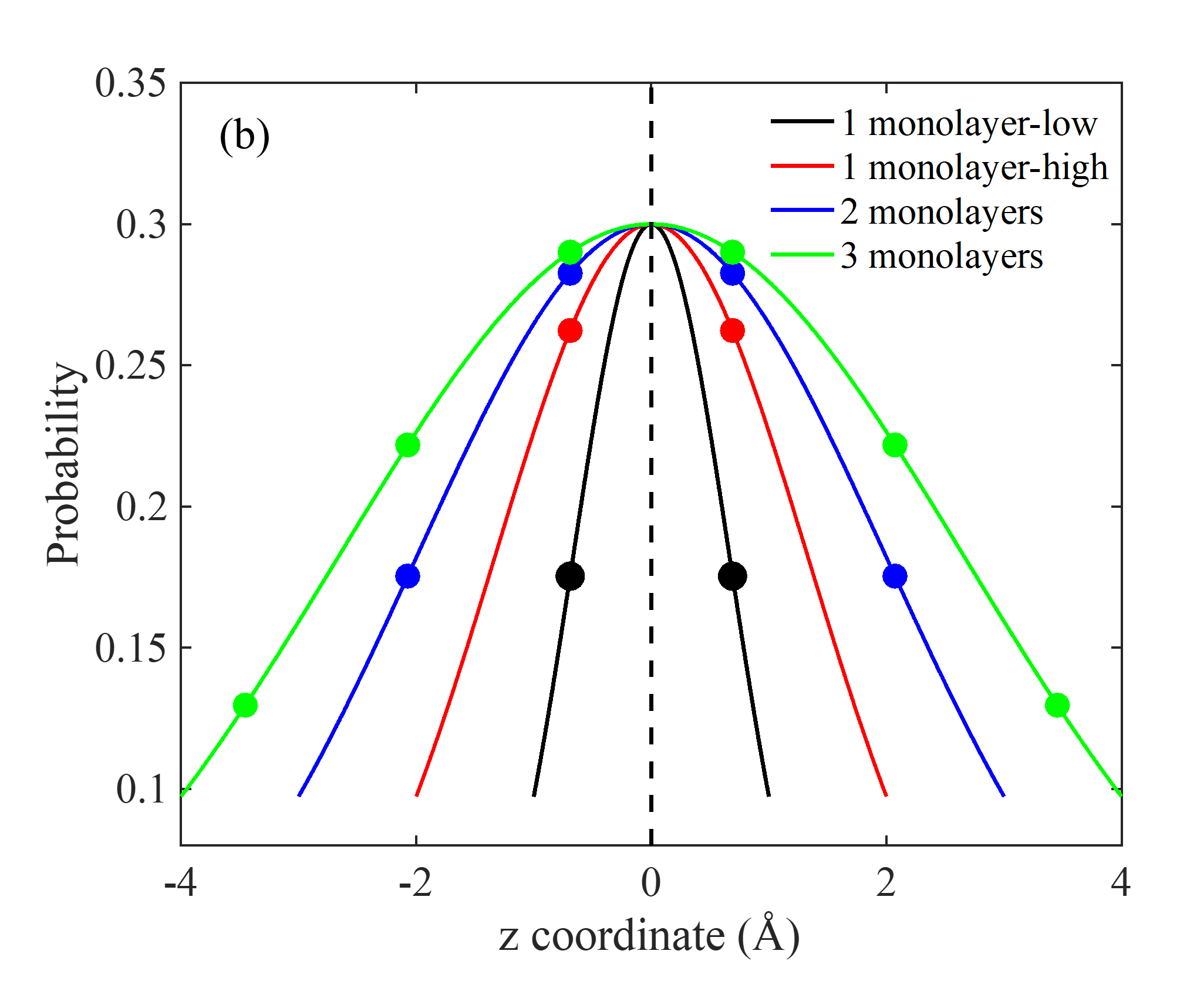}\par
    \caption{Probability of adding vacancies according to Eq.~(1). The markers represent the probability value used for each monolayer of atoms.}
    \label{f-2} 
\end{figure}

\section{Procrustes Analysis}
 
To quantify the structural order of each atom, we apply the statistical shape analysis tool, Procrustes analysis.\cite{Gower} The Procrustes analysis compares the shape of two objects by superimposing one object onto the other and then quantifying the difference between the objects. Procrustes superimposition is performed by optimally translating, rotating/reflecting, and uniformly scaling the objects. In this work, the objects being compared are the local environment of atoms in crystalline Si or Ge (i.e., the reference atoms) and the local environment of each atom in our disordered system (i.e., the studied atoms). Crystalline Si and Ge both have a diamond cubic crystal structure, with each atom bonded to four nearest neighbors in a tetrahedral shape. Since this is the only unique local environment within the unit cell of Si or Ge, the local environment chosen for comparison was simply the atom of interest and its four nearest neighboring atoms. The local environments are extracted by time-averaging the position of atoms in our simulation. Crystalline reference local environments are obtained from atoms in the center cells of the leads, while disordered local environments are from atoms near the interface. 

The process for Procrustes superimposition begins with translation. We translate the reference and studied local environments so that the central atoms are both at the origin. Next, both the reference and the studied local environment are uniformly scaled such that the root mean squared distance of each atom to the origin equals to 1. Then, we determine the optimal rotation transformation that minimizes the root mean squared deviation between corresponding atoms of the two local environments. To do so, the positions of the reference atoms are fixed and singular-value decomposition is applied to determine the optimal rotation matrix transformation to be applied to the local environment. Lastly, the reference and studied local environments are compared by calculating the square root of the sum of the squared distances between the corresponding points (i.e, Procrustes distance, $d$), described by
\begin{equation}
d = \sum_{i=1}^{\rm {\#\ of\ atoms \, = \, 4}}\sqrt{(x_{{\rm {ref}},i}-x_{{\rm {std}},i})^2 + (y_{{\rm {ref}},i}-y_{{\rm {std}},i})^2 + (z_{{\rm {ref}},i}-z_{{\rm {std}},i})^2},
\end{equation}
where the $x_{{\rm {ref}},i}$ and $x_{{\rm {std}},i}$ are the $x$ coordinates of $i$th reference atom and studied atom, and similar for $y$ and $z$. During Procrustes analysis, depending on which the order reference atoms are associated to the studied atoms, the Procrustes distance may be different. To account for this, Procrustes superimposition needs to be completed for all combinations of reference/studied atom pairings, and the smallest Procrustes distance of these combinations is chosen for future analysis. This Procrustes distance is used to quantify the atomic disorder of atoms in our system (i.e., quantified disorder).

The quantified disorder of three structures are plotted in Figure 3, which represents the perfect interface, a slightly defective interface, and  a heavily defective interface. The hollow circles represent the disorder for each individual atom while the red line is the moving average with a window size of 5 Angstroms. Figure 3(a) shows the expected lack of disorder for the perfect interface, with a small effect at the interface. This interface has a continuous structure after relaxation and is highly similar to the bulk structure of Si and Ge. Adding 4 and 18 vacancies creates substantial disorder at the interface in the 1-2 nm range, as shown in Figures 3(b) and 3(c). We observe larger disorder with more vacancies. The Gaussian-like moving average suggests that our designed distribution of vacancies created gradually changing disorder near the interface. 

\begin{figure}
    \includegraphics[width=1\linewidth,trim=10 140 10 140, clip]{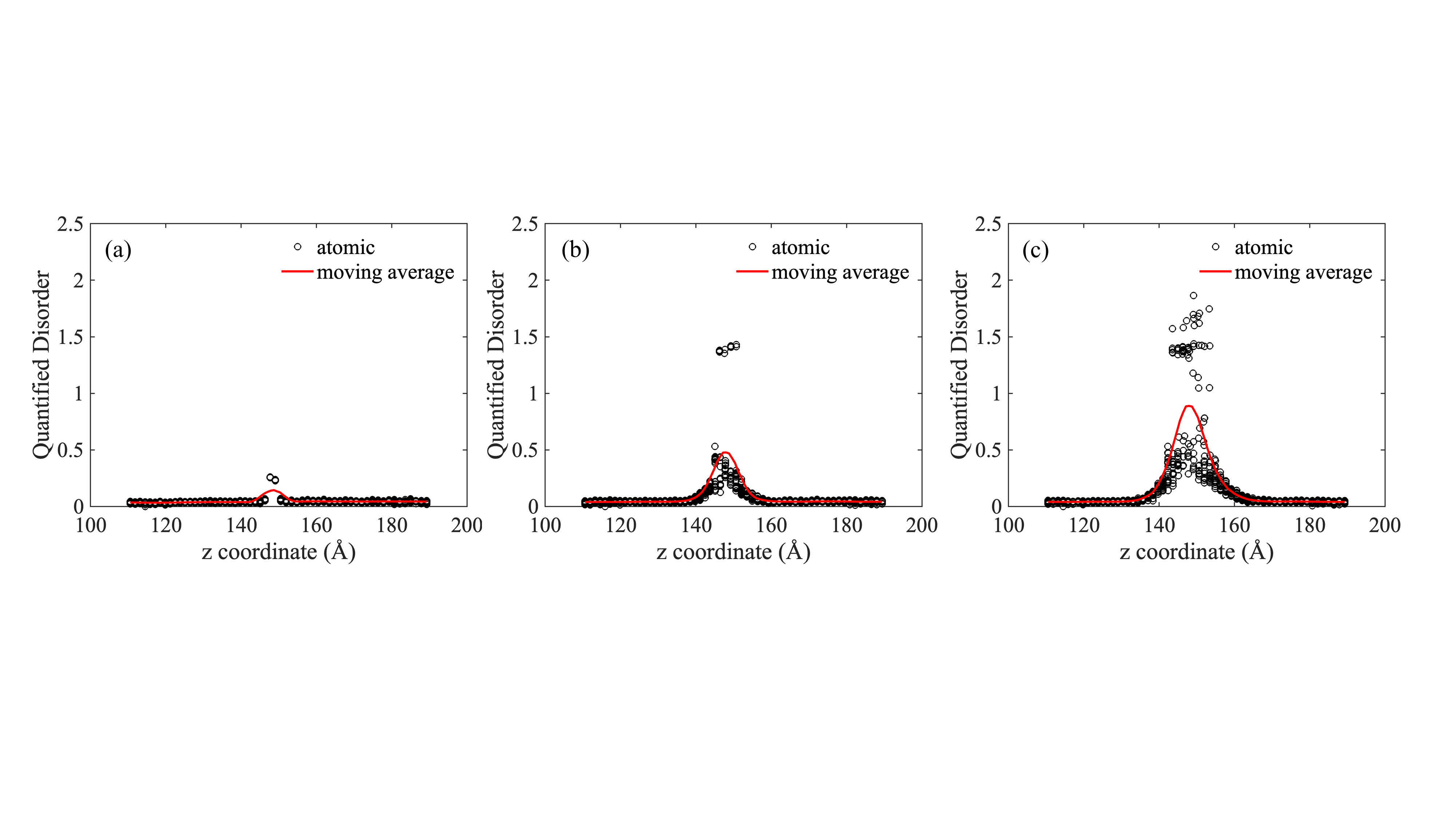}\par
    \caption{Quantified disorder as a function of position within the simulation for (a) perfect interface, (b) 4 vacancies added, and (c) 18 vacancies added. The hollow markers are the quantified disorder for each individual atom in our system. The red line is the moving average quantified disorder.}
    \label{f-5} 
\end{figure}

\medskip

\newpage

\bibliography{main.bib}

\newpage 

\end{document}